# B+-tree Index Optimization by Exploiting Internal Parallelism of Flash-based Solid State Drives

Hongchan Roh[1]    Sanghyun Park[1]    Sungho Kim[1]    Mincheol Shin[1]    Sang-Won Lee[2]

[1]Dept. of Computer Science, Yonsei University    [2] Sungkyunkwan University

{fallsmal, sanghyun, runtodream, smanioso}@cs.yonsei.ac.kr    wonlee@ece.skku.ac.kr

## ABSTRACT
Previous research addressed the potential problems of the hard-disk oriented design of DBMSs of flashSSDs. In this paper, we focus on exploiting potential benefits of flashSSDs. First, we examine the internal parallelism issues of flashSSDs by conducting benchmarks to various flashSSDs. Then, we suggest algorithm-design principles in order to best benefit from the internal parallelism. We present a new I/O request concept, called *psync* I/O that can exploit the internal parallelism of flashSSDs in a single process. Based on these ideas, we introduce B+-tree optimization methods in order to utilize internal parallelism. By integrating the results of these methods, we present a B+-tree variant, PIO B-tree. We confirmed that each optimization method substantially enhances the index performance. Consequently, PIO B-tree enhanced B+-tree's insert performance by a factor of up to 16.3, while improving point-search performance by a factor of 1.2. The range search of PIO B-tree was up to 5 times faster than that of the B+-tree. Moreover, PIO B-tree outperformed other flash-aware indexes in various synthetic workloads. We also confirmed that PIO B-tree outperforms B+-tree in index traces collected inside the Postgresql DBMS with TPC-C benchmark.

## 1. INTRODUCTION
Pioneering studies regarding the features of flashSSDs discovered several key features of flash memory and flashSSDs that are different from those of hard-disks, and focused on addressing DBMS issues with regard to the differences. These DBMS issues involve write-oriented problems such as asymmetric read/write throughput and shortened life-span caused by frequent write-operations. Delta-log based approaches [12, 13], which extract only the updated portions of pages and save them as logs thereby reducing the amount of data to be written, are representative approaches to handle write-oriented problems. Since such previous studies intensively researched on addressing the write-oriented problems, we move our attention to best utilizing advantage of potential benefits of using flashSSDs.

The excellent IOPS performance of flashSSDs is based on their internal parallel architecture. Since flashSSDs embed plural flash memory packages and process multiple I/O requests at the same time, it is possible for a flashSSD to achieve much higher IOPS (Input/Output Operations Per Second) than a flash memory package. However, the outstanding random I/O performance of flashSSDs will remain only a potential performance specification, unless DBMSs take advantage of internal parallelism and fully utilize the high IOPS.

Several flash-aware (flash-memory aware) B+-tree variants have been proposed. The B+-tree index is a good example of how to resolve DBMS issues with regard to flash-based storage devices. Flash-aware B+-tree variants and flash-aware indexes mostly focused on reducing write operations caused by index-insert operations [19, 23] or utilizing sequential pattern benefits [16].

The purpose of this paper is to examine principles of taking advantage of the internal parallelism of flashSSDs, and to optimize the B+-tree index by applying these principles. In this paper, we first present benchmark results on various flashSSDs and known characteristics of the flashSSD parallel architecture, and deduce how to maximize the benefits of internal parallelism. By applying these principles, we introduce new algorithms and a method to determine optimal node sizes of a B+-tree. Eventually, we present a B+-tree variant, PIO B-tree (Parallel I/O B-tree), that integrates the optimization methods into the B+-tree.

This paper is organized as follows. Section 2 describes the internal parallelism and principles to utilize it, with the benchmark results on various flashSSDs. We present B+-tree index optimization method and introduce PIO B-tree as the integrated result of the optimization methods in Section 3. Section 4 describes experimental results, and we list related work in Section 5 and conclude this paper in Section 6.

## 2. INTERNAL PARALLELISM OF SSD
### 2.1 Understandings of Internal Parallelism
Figure 1 presents a flashSSD internal architecture. FlashSSDs are configured with plural flash memory packages. FlashSSDs implement the internal parallelism by adopting multiple channels each of which is connected to a chunk of plural flash memory packages. There exist two types of the internal parallelism such as channel-level parallelism between multiple channels and package-level parallelism between ganged flash memory packages in a chunk [1]. The performance enhancement can be estimated by the factors of channel-level and package-level parallelism.

If there are $m$ channels each connected to a gang of $n$ flash memory packages, the performance gain can be up to $m \times n$ times, compared to the performance of a flash memory package.

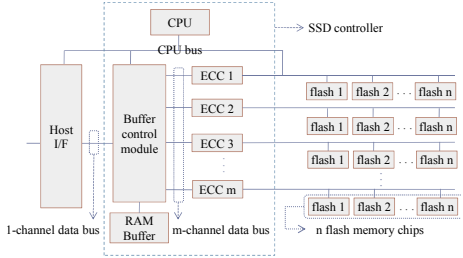

Figure 1. Internal architecture of flashSSDs

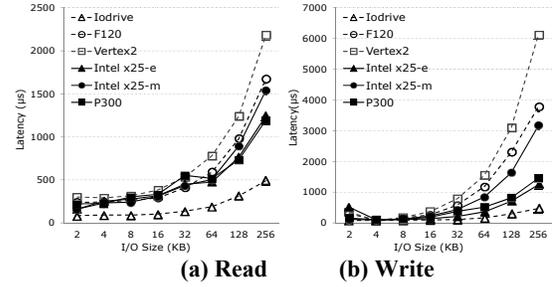

(a) Read  (b) Write

Figure 2. Latencies with different I/O sizes

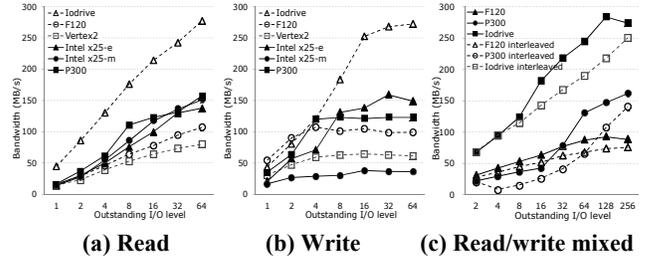

(a) Read  (b) Write  (c) Read/write mixed

Figure 3. Bandwidths with increasing OutStd I/O level

The performance improvement by the channel-level parallelism is somewhat clearer than that of the package-level parallelism. If the host I/F (host interface) requests I/Os designated to different flash memory packages spanning several channels, the channel-level parallelism is achieved by transferring the associated data through the multiple channels at the same time. In this process, command queuing mechanisms (NCQ, TCQ) of the host I/F involve in producing favorable I/O patterns to the channel-level parallelism. The host I/F swaps the queued I/O operations and adjusts the orders of the I/Os in order to make the I/O requests designated to flash memory pages spanning multi channels [14]. Based on the understandings of channel-level parallelism, it is a reasonable inference that the flashSSD performance can be enhanced by requesting multiple I/Os simultaneously.

Package-level parallelism is implemented by striping flash memory packages of each gang. This is analogous to striping a disk array in RAID techniques. The striped flash memory packages in some cases cause a larger logical unit of I/O requests. The striped pages or blocks of the flash memory packages are mostly placed in consecutive LBA (Logical Block Address) regions. In the striped flash memory packages, the write-interleaving technique enhances the write performance by avoiding the shared data-bus channel competition and by interleaving data transfers while other flash-memory packages are programming the already transferred data. This suggests that the SSD performance can be enhanced by requesting I/Os having larger granularity.

We examined the performance effects of these two parallelism factors through benchmark tests on six different flashSSDs. All the benchmarks were conducted in direct I/O mode. We carefully chose these flashSSDs to examine parallelism issues with as many SSD internal architectures as possible. The behaviors of SSDs are determined by three major components, the employed host I/F type, embedded controller, and adopted flash memory. The chosen flashSSDs include the modern host I/F types (SATAII, SATAIII, PCI-E), the controllers of major SSD controller vendors (Intel, Fusion-io, SandForce, Marvell), and flash memory types (SLC 50nm, SLC 35nm, MLC 35nm, and MLC 25nm) [4, 5, 9, 10, 17, 20]. For the tests we used micro-benchmarks, with outstanding I/Os created by using Linux-native asynchronous I/O API (libaio).

For the package-level parallelism tests, we measured the random-read and random-write latency on the flashSSDs, doubling the I/O request size from 2KB at a time. As depicted in Figure 2 (a) and (b), even though the read and write latency increased with respect to the I/O size, the increased pattern was not linear. In several cases, 4KB random-read and random-write latencies were almost the same as or less than 2KB random-read and random-write latencies, which indicates that the bandwidth was enhanced by more than twice. This is because requesting I/Os with large I/O sizes is favorable for the striped flash memory packages. In order to examine the channel-level parallelism, we measured random-read and random-write bandwidths, increasing outstanding I/O level. The outstanding I/O level (OutStd level) indicates how many I/Os are requested at the same time.

Figure 3 (a) and (b) present the benchmark results when read and write operations are separately requested with I/O size fixed at 4KB. The read and write bandwidth was gradually enhanced with increasing OutStd level. We confirmed more than ten-fold bandwidth enhancement by increasing OutStd level in read and write operations both, compared to the read and write bandwidth with the OutStd level of 1, which was the similar to the benchmark results of the previous study [3].

A previous study [3] reported that SSD performance can be degraded with mingled read/write patterns of high outstanding I/Os by the interference between reads and writes. In order to confirm it, we compared the performance of the highly interleaved workload with the nearly non-interleaved workload. Highly interleaved workload was composed of a read operation directly followed by a write operation whereas non-interleaved workload was composed of $n$ number of consecutive reads followed by $n$ number of consecutive writes, in total of $n$ outstanding I/Os at a time, with random I/O patterns in 4GB file. We measured the bandwidth of the two workloads with increasing the OutStd level ($n$) by using micro-benchmarks. Figure 3 (c) shows the result (highly interleaved results were marked with 'interleaved'). The bandwidth of nearly non-interleaved workloads was greater than that of highly interleaved workloads (1.25, 1.37, and 1.3 times greater on F120, P300, and Iodrive at 64 OutStd level).

### 2.2 How to Utilize Internal Parallelism

In order to utilize package-level parallelism, it is required to request I/Os having larger granularities. If I/O requests with larger granularities have less latency, the I/O size can be chosen as a base I/O unit. Otherwise, it is needed to consider trade-offs between increased latency and enhanced bandwidth.

In order to utilize channel-level parallelism, multiple I/O requests should be submitted to flashSSDs at once. Parallel processing is a traditional method to separate a large job into sub-jobs and distribute them into multiple CPUs. Since I/Os of each process can be independently requested to OS at the same time, outstanding I/Os (multiple parallel I/Os) can be delivered to a flashSSD at the same moment.

Due to the high performance gain from channel-level parallelism on a single flashSSD, we need to treat I/O parallelism as a top priority for optimizing I/O performance even in commodity systems, as is stated in [3]. In order to achieve it, more light-weight method is needed since parallel processing (or mutli-threading) cannot be applied in every application programs involving I/Os. In order to best utilize channel-level parallelism, it is also required to deliver the outstanding I/Os to the flashSSDs, minimizing the interval of consecutive I/O requests since inside flashSSDs they can batch-process only the I/O requests gathered in its own request queue (a part of NCQ technology) within a very narrow time span. Therefore, we suggest a new I/O request method, *psync* I/O that creates outstanding I/Os and minimize the interval between consecutive I/O requests within a single process.

## 2.3 Psync I/O: Parallel Synchronous I/O

Two different types of I/O request methods exist in current operating systems. One is synchronous I/O (sync I/O), which waits until the I/O request is completely processed. The other is asynchronous I/O (async I/O), which immediately returns even if the I/O processing is still in progress, thus making it possible for the process to execute next command, but async I/O requires a special routine for handling later notification of I/O completion.

We hope that future OS kernel versions will include system calls for the still conceptual *psync* I/O. *Psync* I/O synchronously operates in the same way as traditional sync I/O except that the unit of operation is an array of I/O requests. We define the three requirements of *psync* I/O as follows. 1) It delivers the set of I/Os to the flashSSDs and retrieves request results at once. Another set of I/O requests can be submitted in sequence only after the results of the previous set are retrieved. 2) The I/Os are requested as a group in the OS user space and the group needs to be sustained until they are delivered to I/O schedulers in the OS kernel space, thereby minimizing the request interval between consecutive I/Os upon I/O schedulers 3) No special routine is required to handle I/O completion events since the process is blocked until the set of I/Os are completely handled.

Since no I/O requesting method that satisfies the *psync* I/O requirements exists in current OSs, we designed a wrapper function that emulates *psync* I/O by using Linux-native asynchronous I/O API. The wrapper function delivers a group of I/O requests to the 'io_submit()' system call by containing them in Linux async I/O data structures (struct iocb), and it waits until all the results are returned, executing 'io_getevents()' system call in Linux. Even though this implementation cannot fully satisfy the requirement 2), this is the best alternative that we found. Likewise, in other operating systems, this alternative method can be also implemented by using their KAIO (Kernelized Asynchronous I/O) APIs.

We compared the performance of the implemented *psync* I/O with the performance of parallel processing by conducting benchmarks in direct I/O mode. In the first benchmark, we measured the bandwidth of each method in a shared file with a mixed read/write setting on the Linux EXT2 file system. *Psync* I/O was configured to request a group of I/Os as many as the OutStd level at a time.

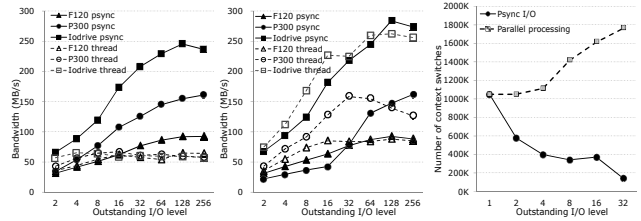

(a) In a shared file (b) In separate files (c) Context switches

Figure 4. Psync I/O and Parallel processing comparison

We compared the result with the bandwidth of parallel processing where each of multi-threads requested a sync I/O and the number of multi-threads was fixed at the OutStd level with a random I/O pattern in a 4GB file. In the second benchmark, we measured the performance with the same settings as the first benchmark except using multiple files (A different file for each thread). Lastly, we measured context switching cost of parallel processing and *psync* I/O as increasing the OutStd level in Linux. As shown in Figure 4 (a), in a shared file, parallel processing performance showed nearly saturated performance at the bandwidth of OutStd level 2. Consequently, *psync* I/O outperformed parallel processing in a shared file. On the contrary, in Figure 4 (b) with separate files, parallel processing showed similar performance to the *psync* I/O performance. The performance degradation in a shared file is because write operations requested in sync I/O coupled with direct I/O cannot be overlapped in a shared file in Linux EXT2 file system. POSIX requires write-ordering for synchronous I/Os, which indicates that writes must be committed to a file in the order in which they are written and that reads must be consistent with the data within the order of any writes. Most POSIX-compliant file systems simply implement a per-file reader-writer lock to satisfy the write-ordering requirement. Therefore, parallel processing cannot utilize the internal parallelism when I/Os are requested into the same file in the file systems. Figure 4 (c) shows the numbers of context switches of each method when 1 million 4KB read requests were given. The context switch count of parallel processing was an order of magnitude greater than *psync* I/O at OutStd level 32. The direct cost and indirect cost of context switching can be much worse in parallel processing than *psync* I/O. For example, a previous study [7] revealed that the time cost caused by context switching is a nontrivial part (nearly 10%) of total cost of DBMS operations. We obtained almost identical results for the three benchmarks by using multi-processes.

We suggest algorithm design principles in order for DBMSs to best benefit from the internal parallelism of flashSSDs based on principles suggested by previous studies [3, 10] and our own findings.

1. *Large granularity of I/Os*: Request I/Os with large granularity in order to utilize package-level parallelism.

2. *High outstanding I/O level*: Create outstanding I/Os in order to utilize the channel-level parallelism. Consider using *psync* I/O first in order to request multiple I/Os in a single process and save parallel processing for later use in more suitable applications (i.e. applications that require both heavy computation and intensive I/Os).

3. *No mingled read/writes:* Avoid creating I/Os in a mingled read/write pattern.

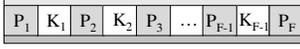

**Figure 5. Internal node structure**

## 3. B+-TREE INDEX OPTIMIZATION
## 3.1 Optimized Algorithms for flashSSDs
In this section, we optimize B+-tree algorithms to utilize channel-level parallelism of flashSSDs, according to *Principle* 2 and 3.

### 3.1.1 Multi Path Search Algorithm
Searches to next-level nodes cannot be performed without obtaining the search results of current-level nodes since index records of the current-level nodes contain the locations of next level nodes.

The only way to achieve *Principle* 2 is to search multiple nodes at the same level. However, this is possible only if a set of search requests are provided at once.

Under the assumption that the set of search requests is given, we design a Multi Path Search (*MPSearch*) algorithm, which processes a set of requests at once while searching multiple nodes level by level. The method to acquire the set of requests is explained later with the detailed algorithm of each index operation.

We represent an internal node of a B+-tree index as a set of key values ($K_i$) and pointer values ($P_i$) each pointing to the location of a child node as depicted in Figure 5, where *F* represents the fanout (the maximum number of pointers).

Let *S* denotes the set of search requests.

$$S = \{s \mid s \text{ is the key value for each search request}\}$$

*S* includes all the key values of search requests, and |*S*| represents the number of given search requests.

The basic concept of *MPSearch* is described as follows.

First, the root node of the B+-tree index is retrieved. The key values of the root node are inspected, and the pointers to the next-level nodes designated by any of the search requests are extracted as (1), where *P* denotes the set of the extracted pointers (refer to 'CheckSearchNeeded' function of Algorithm 1 for more details).

$$I = \{1 \leq i \leq F \mid K_{i-1} \leq s < K_i, s \in S, K_0 = -\infty, K_F = \infty\}$$
$$P = \{P_i \mid i \in I\} \quad (1)$$

Second, the next-level nodes designated by the pointer set (*P*) are read at once through *psync* I/O. The entries of the read internal nodes are examined node by node, and the pointer set for each node corresponding to *S* is extracted by using (1). The extracted pointer sets create an array of the pointer sets *P′* as follows.

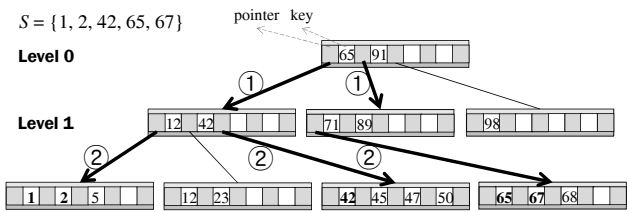

**Figure 6. MPSearch with two psync I/Os**

$$P' = \{P \mid P \text{ for each internal node }\}$$

Third, all the child nodes designated by the pointers in the pointer sets of *P′* are read at once through *psync* I/O. The entries of the read internal nodes are examined for every node, and for each node the pointer set to the next-level nodes is extracted, creating an array of the pointer sets *P′*, again. This process is repeated, until it reaches leaf nodes and retrieves all the leaf nodes corresponding to the search requests *S*.

Figure 6 presents an example of *MPSearch* when the number of search requests is 5. With 2 *psync* I/O calls, the leaf nodes having the search keys are retrieved by *MPSearch*.

During this procedure, *psync* I/O is executed (*treeHeight*-1) times, at maximum processing |*S*| read requests for each time, since *psync* I/O is executed once for every level except the root level.

This indicates that MPSearch can achieve |*S*| OutStd level at maximum. It also implies that |*S*| in-memory buffer pages are required for every *psync* I/O call since one buffer-page is needed for every node to be loaded into main memory. This might consume a significant amount of main memory space if |*S*| is considerably large.

Therefore, we adopt a parameter called *PioMax* that indicates the maximum number of I/Os submitted to a *psync* I/O call. By doing so, the maximum main memory space is limited to $PioMax \cdot (treeHeight - 1)$ pages. As demonstrated in the results of Figure 3, *PioMax* is not necessary to be considerably large. A moderate value (around 32) can increase the bandwidth enough.

*MPSearch* process needs to be adjusted reflecting this change.

---

**Algorithm 1**: Multi Path Search

**Procedure MPSearch**(S[], P[], PioMax, PioCnt, level, b[][])
**Input:** S[] (search keys), P[] (pointers to target nodes), *PioMax* (max number of I/Os at a time), PioCnt (number of I/O requests to psync I/O), tree level, b[][] (2$^{nd}$ dim array of buffer pages, b [i][j]: j$_{th}$ buffer page on i$_{th}$ level), F (node fanout).
**Output:** leafNode[] (leaf nodes corresponding to S[])

1: cnt := 0
2: **if** (level = (*treeHeight*-1)) **then**   //leaf nodes
3:   b[level][] := psync_read(P[])
4:   leafNode[] := b[level][]   //convert buffers to leaf nodes
5:   **return** leafNode[]
6: **else**   //non leaf-nodes
7:   b[level][] := psync_read (P[]), isEnd = false
8:   **for** n := 0 **to** PioCnt-1 **step** 1 **do**
9:     node := b[level][n]   //covert buffer to node structure
10:    **if** (n = PioCnt-1 and i = F) **then** isEnd = true
11:    **for** i := 1 **to** F **step** 1 **do**
12:      **if** (**CheckSearchNeeded**(i, node.K[], S[])) **then**
13:        P[cnt++] := node.P$_i$ //P$_i$ is i$^{th}$ pointer of the node
14:        **if** ((cnt ≠ 0 and (cnt % PioMax) = 0) or isEnd) **then**
15:          **MPSearch**(S[], P[], PioMax, cnt, level+1, b[][])
16:          Reset P[], cnt := 0

**function** CheckSearchNeeded(index, K[],S[])

17: K[0] := −∞, K[*F*] := ∞   //K[i] is i$^{th}$ key value ($K_i$) of a node
18: **for** i := 1 **to** |S| **step** 1 **do**
19:   **if** (K[index-1] ≤ S[i] < K[index]) **then**
20:     **return** true
21: **return** false

Algorithm 1 presents the complete process of *MPSearch* with *PioMax* considered. *MPSearch* begins with reading the root node (line 7). Then the pointers to the child nodes relevant to the set of search keys are extracted, and the relevant pointers create the pointer set (line 12-13, line 17-21). Each pointer set has a size less than *PioMax*, and in turn the pointer set is delivered to *MPSearch* for the sub-tree traverse (line 13-15). If the number of relevant pointers is greater than *PioMax*, then the procedure recursively calls the procedure itself only with the first pointer-set having pointers as many as *PioMax*, leaving the job to handle the rest of the pointers to a later task (line 14-15). After the sub-tree traverse by the recursive call is finished, the remaining pointers are processed by later recursive calls. This part is analogous to the Depth First Search (DFS).

In the recursive call from the root node, internal nodes on level 1 are read via *psync* I/O (line 7). The *psync* I/O call retrieves the requested nodes at once from the locations provided by the given pointer set P[], and loads them into the given buffers b[level][]. In the retrieved node, the procedure extracts the relevant child-node pointers (line 12, 17-21), and creates a pointer set for every *PioMax* pointers (line 8-13). For each pointer set, the procedure is recursively called (line 15). When this procedure reaches the leaf level ($treeHeight - 1$), the procedure retrieves the leaf nodes designated by the given pointer set via *psync* I/O (line 2-5).

### 3.1.2 Parallel Range Search
It is straightforward to apply *MPSearch* into the range search. This is simply achieved by requesting the search request set $S$ defined as (2) via *MPSearch*.

$$S = \{s \mid range.start \leq s < range.end\} \qquad (2)$$

The *MPSearch* retrieves the leaf nodes including the entries with key values in the range.

The traditional method to conduct a range search is reading the leaf nodes that are linked between each other one by one in sequence, after searching the first leaf node containing an entry with the least key value of the range. The new range search, called parallel range (*prange*) search reads relevant internal nodes level by level via *psync* I/O until it reaches to the leaf level.

*Prange* search reads more internal nodes than the traditional range search. Nevertheless, *prange* search outperforms the legacy range search in general since the benefit from leaf-node reads by using *psync* I/O is substantial (up to ten-fold bandwidth increase). We discovered that *prange* search time was always less than or equal to the legacy range search time in any condition on the flashSSDs tested in this paper (see the empirical study in Section 4.1.2).

### 3.1.3 Update Operations
In this section, we optimize update operations such as insert, delete, and update by utilizing the channel-level parallelism of flashSSDs. Hereinafter an update operation indicates an update-type operation including an index-insert, index-delete, and index-update operation unless we further differentiate it as an index-update operation.

The index records of update operations are inserted in an in-memory structure (Operation Queue) to accumulate a group of update operations for later *MPSearch-like* batch-updates. Each update operation is completed immediately after its index-record is inserted into the Operation Queue (OPQ) as an OPQ entry. OPQ entries are not written to the flashSSDs until the OPQ entries are batch processed. Since update operations are not immediately reflected to flashSSDs and reside on the in-memory structure for a while, this method requires additional features to the traditional DBMS recovery scheme for avoiding data-loss during system crashes. We address this issue in Section 3.4.

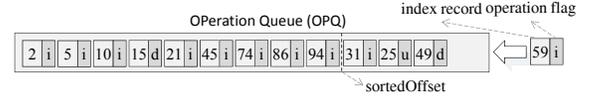

**Figure 7. Operation Queue structure**

**Operation Queue:** We first describe OPQ that provides an in-memory space for index records of update operations. The update operations reside in OPQ, until they are processed by a batch-update operation (*bupdate*). The adoption of OPQ requires search algorithms to first scan the entries in OPQ before traversing trees so that the entries of queued update operations can be successfully searched.

OPQ is an array-based structure, including index records of the queued update operations in its elements called OPQ entries.

- OPQ Entry (Ent): consists of an index record and an operation flag that indicates the type of the update operation.
    - Ent.indexRec: an index record containing the key value and pointer to the data record page (data page id).
    - Ent.op: an operation flag indicating the update operation type (i: insert, d: delete, u: update)

Figure 7 shows an example of OPQ. The array region is divided into two parts, one for the sorted array region and the other for the recently appended entries. The two regions are differentiated by *sortedOffset*. We configured OPQ in this manner to consider trade-offs between the in-OPQ search cost for searching index records within OPQ and the OPQ append cost for inserting an entry to OPQ. For every update operation, OPQ creates an OPQ entry and merely appends it into the next slot of the most recently appended entry without considering the orders between key values. The OPQ append cost is minimal since only one main memory page is accessed for an update operation. The sorting occurs on the basis of a parameter called sort period (*speriod*). For every *speriod* update operations, a sort operation for OPQ entries is executed. Since there are already sorted entries in the sorted region (before *sortedOffset*), there is no need to sort the entire region. The recently appended entries in the unsorted region (after *sortedOffset*) are first sorted. Next, they are merged into the entries of the sorted region. The merge process is analogous to that of the merge-sort algorithm. Due to these features, in-OPQ searches can be conducted by using binary search in the sorted region, leaving the unsorted entries to the linear search.

**Batch Update:** By using OPQ, *MPSearch* can be applied to batch-processing of the queued update operations. First the search request set is defined as the following.

$$S = \{s \mid s \text{ is the key value for each update operation in OPQ}\}$$

The batch-update operation (*bupdate*) is triggered by the event when OPQ is fully filled with the update operations. Before the procedure begins, OPQ entries are sorted by the aforementioned process. Until the *bupdate* procedure reaches leaf nodes, the *bupdate* process is the same as *MPSearch*. After reading multiple leaf nodes via *psync* I/O, the update operations of OPQ are performed to the leaf nodes, and the updated leaf nodes are written to flashSSDs at once via *psync* I/O. In addition, differently from the original update operation, multiple fence keys can be generated by multiple node-splits, causing propagation of the fence keys to a parent node.

**Algorithm 2**: Batch Update
**Procedure bupdate**(U[], P[], PioMax, PioCnt, level, b[][])
**Input:** U[] (OPQ entries), P[] (pointers to the target nodes), PioMax (max number of I/Os at a time), PioCnt (number of I/O requests to psync I/O), tree level, b[][] ($2^{nd}$ dim array of buffer-pages, b [i][j]: $j_{th}$ buffer-page on $i_{th}$ level), F (fanout).
**Output:** $K_f$[] (OPQ entries with fence key index records)
1: cnt := 0
2: **if** (level = (*treeHeight* -1)) **then**   //leaf nodes
3:   b[level][] := psync_read(P[])
4:   leafNode[] := b[level][]   //covert buffers to the leaf nodes
5:   **for** n := 0 **to** PioCnt-1 **step** 1 **do**
6:     $K_f$[n]:= **updateNode**(leafNode[n], U[], level, n, b[][])
7:   Reset P[] with the pointers of the updated leaf nodes
8:   psync_write(P[], b[level][])
9:   **return** $K_f$[]
10:**else**   //non leaf-nodes
11:   b[level][] := psync_read (P[]), isEnd := false
12:   node[] := b[level][]   //covert buffers to the internal nodes
13:   **for** n := 0 **to** PioCnt-1 **step** 1 **do**
14:     **for** i := 1 **to** F **step** 1 **do**
15:       node := node[n]
16:       **if** (n = PioCnt-1 and i = F) **then** isEnd = true
17:       **if** (**CheckSearchNeeded**(i, node.K[], U[])) **then**
18:         P[cnt++] := node.$P_i$ //$P_i$ is $i^{th}$ pointer of the node
19:         **if** ((cnt ≠ 0 and (cnt % PioMax) = 0) or isEnd) **then**
20:           U'[]:={u ∈ U[] | node.K[i-1] ≤ u.key < node.K[i]}
21:           $K_f$'[]:=**bupdate**(U'[],P[],PioMax,cnt,level+1,b[][])
22:           Append $K_f$'[] to $K_f$[], and Reset P[]
23:           cnt := 0
24:     $K_f$[n]:=**updateNode**(node, $K_f$[], level, n, b[][])
25:   Reset P[] with the pointers the updated internal nodes
26:   psync_write(P[], b[level][])
27:   **return** $K_f$[]
**Function updateNode**(node, U[], level, n, b[][])
28: Execute each index operation in U[]
29:**if** node is fully filled **then**
30:   Perform node split and get the fence key
31:   $K_f$.indexRec := the fence key and new leaf node pointer
32:   $K_f$.op := 'i'   //insert the fence key record to parent node
33:**else if** node became underflow **then**
34:   Perform node redistribution or node merge
35:   $K_f$.indexRec := the fence key and pointer
36:   $K_f$.op := 'u'   //update the fence key record to parent node
37:   **if** node merge occurred **then** $K_f$[n].op := 'd'
38: b[level][n] := the updated node
39:**return** $K_f$

Algorithm 2 presents the detailed process of *bupdate*. First, it traverses nodes analogous to *MPSearch*, until it reaches leaf nodes (line 10-21). After reading leaf nodes corresponding to *S*, the procedure performs the index operations contained in the OPQ entries to the leaf nodes (line 3-6, 28-39). If a leaf node is fully filled with inserted index records, a split operation is performed (line 29-30). If the leaf node becomes underflow by index-delete operations, a redistribution or merge operation is performed (line 33-34). There is no difference between the detailed process of these operations (node-split, redistribution, and node-merge) from the original ones except that they are batch-processed. The updated leaf nodes are written to the flashSSDs via *psync* I/O (line 8). When the multiple nodes are split, the multiple fence-key records are required to be propagated to the parent nodes (line 6, 31-32). After receiving the propagated fence-key records from leaf nodes, the *bupdate* procedure updates the internal nodes with the fence keys (line 21-24). Likewise, the propagated fence-key records by node-merge and redistribution are handled. Since the fence-key records are inserted to internal nodes, split operations on internal nodes can also take place, causing another fence-key record propagation to their parent nodes (line 24, 29-32). Likewise, merge or redistribution operations are performed as well. This propagation process can be repeated, traversing backward to the root node. During the process, the updated internal nodes by the propagated fence-key records are written to flashSSDs via *psync* I/O (line 26).

Since *bupdate* takes advantage of internal parallelism by using *psync* I/O for read/write operations, the total time for the execution of the update operations can be considerably reduced. In the aspect of latencies, most update operations are instantly finished by merely being appended to OPQ. However, the latency of a certain update operation that triggers the *bupdate* procedure is lengthened until the entire *bupdate* process is completed. This is an acceptable compromise considering the lessened amortized cost for each update operation. Nevertheless, we provide a method to reduce the latency by allowing the users to provide a parameter, called a batch count (*bcnt*). The batch update procedure chooses a group of OPQ entries as many as *bcnt* from OPQ, and processes the operations only. Whenever the *bupdate* process terminates, the chosen entries are eliminated from OPQ. This method alleviates the latency of the update operation initiating *bupdate* procedure.

## 3.2 B+-tree Node Size Optimization
In this section, we focus on enhancing B+-tree performance by utilizing package-level parallelism according to *Principle* 1. We first present a new method to determine the optimal node size and introduce methods to alleviate the cost of enlarged leaf nodes.

### 3.2.1 B+-tree Optimal Node Size on flashSSDs
In order to utilize package-level parallelism, larger I/O size is desirable. Therefore, we considered enlarging the node size of B+-tree according to *Principle* 1. Even though enlarging the node size enhances flashSSD bandwidths, it also increases the latency of each I/O operation. Since this induces search performance degradation, we cannot simply enlarge the node size. Enlarged nodes, however, shortens the B+-tree height. Therefore trade-offs exist between the increased latency and shortened height.

Considering the trade-off, Graefe [6] presented the optimal B+-tree node size (2KB node size) for flash memory. The optimal node size is determined by the utility/cost method defined as (3).

$$IndexPageUtility / IndexPageAccessCost \quad (3)$$

*IndexPageUtility* is defined as $\log_2(EntriesPerPage)$, which represents the factor to reduce the B+-tree height, meanwhile *IndexPageAccessCost* is defined as the time to read a node.

The optimal node size (2KB) determined by the utility/cost method and method itself are valid on raw flash memory. Since the read/write costs (latencies) of a node linearly increases with respect to the node size, the size of the smallest I/O unit (a flash memory page, usually 2KB) is the best node size on flash memory. This is no longer true for flashSSDs, however, since read/write latencies of a node are lengthened nonlinearly with increasing node size on flashSSDs. Thus, a more delicate method is needed in order to determine the optimal B+-tree node size on flashSSDs.

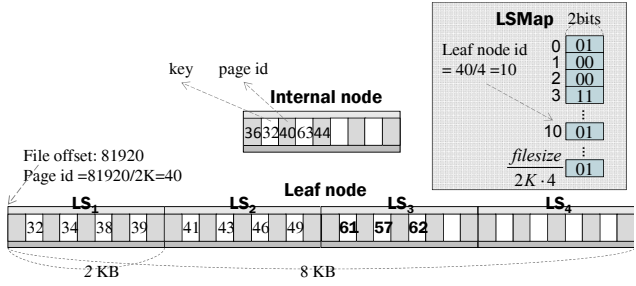

**Figure 8. Leaf node schematic with 2KB page size and leaf node size 4**

The read/write operation has to be separately considered. Since on hard-disks read/write latencies are the same, there is no need to separately consider the read/write operation. On raw flash memory, since read/write latencies linearly increase, the optimal node size is the smallest I/O unit regardless of asymmetric read/write latencies.

However, FlashSSDs not only have the asymmetric read/write latency but also have the feature that the read/write latencies are not linearly increased as the I/O size grows. Therefore, the utility/cost measure has to be extended for the insert/search ratio to be considered on flashSSDs.

We propose a new B+-tree cost model and the method to determine the optimal node size based on the cost model. We first define the cost model by using the notations presented in Table 1. The average latency of a search and an insert operation can be represented as $H \cdot P_r$, and $(H \cdot P_r + P_w)$, respectively. The average latency of an operation regardless of the operation type can be represented as $C_{b+}$ in (5). Index-delete/index-update operations are not considered separately since each operation cost is almost the same as an index-insert operation cost. For simpler estimation, we ignored the amortized leaf node split cost per index-insert operation.

$$H = \log_2 N / \log_2 F' \quad (4)$$

$$C_{b+} = R_s \cdot (H \cdot P_r) + R_i \cdot (H \cdot P_r + P_w)$$
$$C_{b+} = \log_2 N / \log_2 F' \cdot P_r + R_i \cdot P_w \quad (5)$$

The cost gap caused by workload configuration worsens when the buffer manager is employed. This is because node write cost can have more influence on the average response time if the buffer pool lessens the node read cost by caching nodes in main memory.

We present another cost model (6) to consider the buffer pool size. The detailed process to derive this is described in Appendix.

$$C'_{b+} = \left( \lfloor \eta \rfloor + \left( 1 - 1/F'^{(\eta \% 1)} \right) \right) \cdot P_r + R_i \cdot P_w$$

, where $\eta = \log_{F'} \frac{N}{M} - 1 \quad (6)$

With the buffer manager considered, the optimal node size $S_{opt}$ on flashSSDs is determined by $S_{opt} := \arg\min_{size}(C'_{b+})$.

### 3.2.2 Asymmetric Leaf Node

Here, we introduce methods to alleviate the cost caused by enlarged leaf nodes. First, we define the leaf node resizing unit, called Leaf Segment (LS), to vary the leaf node size. In our approach, the leaf node size is only changed, thus making the leaf node size asymmetric with the internal node size.

**Leaf Segment (LS):** Figure 8 presents a leaf node represented with 4 LSs. The leaf node size scales from the size of an LS to multiples of the LS size. Hereinafter, the leaf node size is represented in terms of the number of LSs consisting of the leaf node (e.g. the leaf node size of Figure 8 is 4). The size of an LS is the same as that of a page and an internal node.

**Append-only feature:** We relax a constraint that entries in every B+-tree node have to be sorted in the ascending order of the key values. This constraint makes on average a half of the entire leaf node updated for every index-insert operation. In contrast, our approach is to append an index record right next to the most recently inserted entry in the type of an OPQ entry, regardless of the order between the key values. This makes only one LS updated, thereby reducing the average page-write cost caused by an insert operation from $L/2$ pages to 1 page. Likewise, index records of index-update and index-delete operations are also appended to the leaf node in the type of OPQ entries. Due to the append-only feature, the read cost of a leaf node for an update operation can be reduced from reading the entire leaf node to reading a half of the leaf node. This is because the last LS always exists in the back half part due to the half-full feature of B+-tree nodes. We further reduce the read cost by using an in-memory bitmap structure (LSMap), caching the ID number of the last LS (*last LS ID*), for every leaf node. When *last LS ID* is stored into the LSMap, the ID number is subtracted by $\lfloor L/2 \rfloor$. In contrast, it is added by $\lfloor L/2 \rfloor$ when it is retrieved from the LSMap.

**Shrink Operation:** When leaf nodes are fully filled with appended OPQ entries, after reading the entire leaf nodes, the index-delete or index-update operations are first performed, thereby reducing the number of OPQ entries in the leaf node. This process is called a *shrink* operation. The *shrink* operation is achieved by canceling index-insert operations with the index-delete operations having the same index records. Index-update operations are similarly handled, considered as a sequence of an index-delete and index-insert operation. After the *shrink* operation, if the leaf nodes are still fully filled with entries, node-split operations are performed. If the leaf node becomes underflow by the *shrink* operation, a redistribution or merge operation is performed. In these operations, the entries in the leaf node are sorted and then divided into two nodes or merged into one node.

**Table 1. Notations**

| Notation | Description |
|---|---|
| $H$ | height of a B+-tree index |
| $F$ | maximum number of pointers in an internal node |
| $N$ | current total number of inserted entries |
| $U$ | average node utilization ratio |
| $F'$ | average number of entries in a node with the node utilization ratio. (i.e. $(F - 1) \cdot U$) |
| $P_r$ | random read latency of a page |
| $P_w$ | random write latency of a page |
| $L$ | leaf node size (number of pages) |
| $\ell$ | level of a node |
| $P_r(L)$ | random read latency of a leaf node with size $L$ |
| $R_i$ | insert ratio of the given workload |
| $R_s$ | search ratio of the given workload |
| $M$ | available main memory space in terms of the number of pages |
| $O$ | OPQ size in terms of the number of pages |
| $P'_r$ | amortized response time for a page via *psync* I/O |
| $P'_w$ | amortized response time for a page via *psync* I/O |

## 3.3 PIO B-tree

Thus far, we introduced B+-tree index optimization methods. Here, we present the B+-tree variant, called PIO B-tree, which is an integrated result of the optimization methods. Due to the integrating process, a few differences are made from the original algorithms of the optimization methods.

First, since PIO B-tree has the asymmetric leaf-node structure, *bupdate* request less I/Os. Only one page (the last LS) is read and written on the leaf-node level for every update operation via *psync* I/O. Update operations are appended to leaf nodes in the type of OPQ entries, later inducing the *shrink* operation. In order to reflect these changes, we present the adjusted updateNode function of *bupdate* in Algorithm 3.

Second, *prange* search and point-search algorithm include one more step to inspect the entries in OPQ. Prior to traversing the nodes, the search procedures inspect if there are update operations with the key values they are looking for. The retrieved OPQ entries from leaf nodes are inspected along with the retrieved entries from OPQ. The insert-type OPQ entries are canceled by the delete-type OPQ entries having the same index records. If no insert-type OPQ entry exists, then the failure flag is returned. Otherwise the success flag is returned.

## 3.4 Crash Recovery of PIO B-tree

The write-ahead logging (WAL), steal and no-force buffer management, and multi–phase recovery procedure are common techniques for transaction support in most traditional DBMSs [18]. The state of the art in recovery methods is best illustrated by the ARIES [18] recovery method, where the recovery procedure consists of three phases such as the analysis phase, redo phase, and undo phase.

There is a critical issue of crash recovery with the PIO B-tree's OPQ structure. 1) Since OPQ structure contains the index records of update operations as OPQ entries in volatile memory, data-loss will occur during system crashes. 2) The OPQ structure does not flush the OPQ entries onto flashSSDs in the FIFO (First In First Out) fashion, and thus the conflicting order of index operations cannot be preserved, causing an inconsistent database.

It is worth noting that the OPQ can be regarded as a compacted version of a buffer cache where each OPQ entry corresponds to a dirtied buffer in a DBMS buffer cache. OPQ is a different form of write-back cache that contains not the entire dirtied buffer page (an index node) but only a dirtied element (an index record). In this perspective, OPQ can be regarded as a different type of buffer cache with no-force buffer management policy without WAL. Even in the traditional DBMSs, without logging, no-force buffer management can also cause data-loss and inconsistent database after system crashes. With the support of WAL and redo recovery, data-loss can be prevented, and the conflicting orders can be preserved. If we provide a method to generate transaction logs for each OPQ entry and apply WAL in managing the logs, then the two problems can be resolved.

**Table 2. PIO B-tree specific Transaction Log**

| Log Type | Example ($T_i$: transaction ID, $R_i$: index relation ID) |
|---|---|
| *logical redo log* | <$T_i$, $R_i$, Operation-type, Index record> |
| *flush event log* | <$T_i$, $R_i$, Flush Start, Key range>, <$T_i$, $R_i$, Flush End, Key range> |
| *flush undo log* | <$R_i$, Index node ID, Undo info> |

**Algorithm 3**: Modified updateNode function
**Function updateNode**(node, U[], level, n, b[][])
1: **if** node is an internal node **then**
2:     Update node with U[]
3: **else**
4:     Append U[] to the last LS of the node
5:     if node is fully filled then
6:        shrink(node)
7: **if** node is fully filled **then**
8:     Perform node split and get the fence key
9:     $K_f$.indexRec := fence key and new leaf node pointer
10:    $K_f$.op := 'i'  //insert the fence key record to parent node
11: **else if** node became underflow **then**
12:     Perform node redistribution or node merge
13:     $K_f$.indexRec := the fence key and pointer
14:     $K_f$.op := 'u'  //update the fence key record to parent node
15:     **if** node merge occurred **then** $K_f[n]$.op := 'd'
16: b[level][n] = the updated node
17: **return** $K_f$

The traditional way to generate transaction logs is to extract the updated portion of the dirtied buffer page as a physiological log [18]. Extracting physiological logs cannot be applied to PIO B-tree since PIO B-tree retrieves no buffer for update operations but only append an OPQ entry to the OPQ. However, a *logical redo log* (see Table 2) corresponds to an OPQ entry can be created for every OPQ append operation. Using these logical logs, the corresponding logical redo operations can be performed in the redo recovery phase. An OPQ flush operation can be also regarded as a batch process version of dirty buffer write operations in the traditional buffer management. Therefore, WAL for PIO B-tree must satisfy the following two conditions. First, before index operations are committed, the corresponding logical logs have to be written to flashSSDs. Second, all the logical logs of the OPQ entries chosen for the OPQ flush operation have to be written onto flashSSDs, prior to the OPQ flush operation. Only after the logical log writes are completed, the OPQ flush operation can proceed. In addition, when OPQ flush operations are performed, all the updated nodes during the flush operations are written to flashSSDs (write-through). PIO B-tree also flushes all the OPQ entries in the OPQ and makes it empty when the DBMS system needs to checkpoint.

In this way, PIO B-tree can recover the in-memory OPQ entries lost during the system crash and preserve the conflicting orders. However, still several concerns remain. First, the OPQ flush operation needs to be atomic since system crash during an OPQ flush operation can cause an inconsistent database. Second, logical redo operations on already flushed OPQ entries need to be prohibited since logical redo operations are usually not idempotent. Third, uncommitted index operations should be rolled back. For the three requirements, we suggest the following solutions. First, for guaranteeing the atomicity when an OPQ flush operation is started and finished, the special transaction logs, a pair of *flush event logs*, are written, and for the each update to an index node, a *flush undo log* is generated (see Table 2). Therefore, later in recovery process, if any incomplete flush operation is found by inspecting the *flush event log* pair, then undo operations corresponding to the *flush undo log*s are performed. In recovery process, these *flush undo* operations are first performed prior to the redo phase. In the redo phase, only the OPQ entries that have never been flushed are redo-processed by inspecting key ranges of completed flush operations written in *flush event logs*. The logical redo logs included in the key ranges are ignored in the redo phase. Lastly, for uncommitted index records, PIO B-tree flushes no OPQ entries of uncommitted updates (no-steal policy), so nothing needs to be performed in the undo phase of the recovery process.

## 3.5 Cost Analysis

We provide cost estimation for PIO B-tree. This estimation is for the case when no wide range search is requested. The average cost for an index operation can be represented as (7) based on (5).

$$C_{pio} = R_s \cdot Search + R_i \cdot Insert$$

, where $Search = (H-1) \cdot P_r + P_r(L)$,

$$Insert = \sum_{\ell=0}^{H-2} \left(\frac{1}{G(\ell)}\right) \cdot P_r' + \frac{(P_r' + P_w')}{G(H-1)} \quad (7)$$

$$G(\ell) = \frac{\text{\# of OPQ entries}}{\text{\# of level } \ell \text{ nodes}} = \frac{O \cdot F'/U}{N/\left(F'^{H-\ell} \cdot L\right)}, \text{ and } 1 \leq G(\ell) \leq bcnt \quad (8)$$

The major difference from (5) is that the node-search cost of a search operation in PIO B-tree is different from the node-search cost of a search operation in B+-tree. Since the leaf node size can be asymmetric with the internal node size in PIO B-tree, the cost for reading a leaf node $P_r(L)$ varies depending on $L$ in a search operation. Another difference from (5) is that the node-search cost of an update operation is different from that of a search operation. In the *bupdate* process, the procedure read only once the same internal node requested multiple times by update operations. The function (8) presents the average number of the update operations to read the same node depending on the node level ($\ell$). It is estimated by dividing the total number of OPQ entries with the total number of the nodes in the level, and its value ranges from 1 to *bcnt*. Since for the multiple update operations the normal update process reads multiple times the same node but the *bupdate* process only once reads the node, the average cost decreases by the factor of $G(\ell)$. Furthermore, in *bupdate* process, the node-read and node-write operation is conducted via *psync* I/O, and thus the read and write cost becomes $P_r'$ and $P_w'$, which is considerably less than $P_r$ and $P_w$, respectively. As can be expected from (8), PIO B-tree not only enhances the update performance by using *psync* I/O but also reduces the number of I/Os by eliminating duplicated reads and writes to the same nodes in multiple update operations.

With the buffer pool considered, the average cost of PIO B-tree can be represented as $C_{pio}'$ (refer to Appendix for the process to derive it).

$$C_{pio}' = R_s \cdot Search' + R_i \cdot Insert'$$

, where $Search' = \left(\lfloor \eta \rfloor - 1/F'^{(\eta \% 1)}\right) \cdot P_r + P_r(L)$,

$$Insert' = \left(\sum_{\ell=\lfloor \eta \rfloor}^{H-2} \left(\frac{1}{G(\ell)}\right) - \frac{1/F'^{(\eta \% 1)}}{G(\log_{F'}(M-O)-1)}\right) \cdot P_r' + \frac{(P_r' + P_w')}{G(H-1)}$$

$$\eta = \log_{F'} \frac{N}{L \cdot (M-O)} - 1 \quad (9)$$

## 3.6 How to Choose Leaf Node and OPQ size

When insert ratio is not extremely high, the leaf node size can be set to 4KB to 16KB since the performance of most flashSSDs is optimized with around 8KB I/O size as shown in Figure 2. It is not needed to allocate large main memory space to OPQ, since only with the OPQ size of one (4KB), PIO B-tree demonstrates outstanding performance in update operations as is empirically verified in the experiments (up to 8.2 times faster than B+-tree, see Section 4.1.3). The leaf node size and OPQ size can be determined based on this knowledge. Furthermore, PIO B-tree is able to automatically tune itself by using (10) when it is initially built in order to obtain the best attainable performance.

$$(L_{opt}, O_{opt}) := \arg \min_{L,O} C_{pio}' \quad (10)$$

With the given insert ratio and search ratio, PIO B-tree conducts a micro-benchmark to obtain the flashSSD specifications such as $P_r$, $P_w$, $P_r(L)$, $P_r'$, and $P_w'$, when the PIO B-tree index is initially built. By using the values, the PIO B-tree index determines the optimal leaf node size and OPQ size on the basis of (10). Finally, the PIO B-tree is constructed by using the determined $L_{opt}$, and $O_{opt}$.

## 4. EXPERIMENTAL RESULTS

In this section, we present the experimental results of PIO B-tree. First, we evaluate the effectiveness of the proposed optimization methods by comparing the performance of PIO B-tree with B+-tree. Then, we compare PIO B-tree with other flash-aware indexes such as BFTL [23], and FD-tree [16] in synthetic workloads. The synthetic workloads were used for a better control of the workload configuration. For the evaluation in a more realistic configuration, we generated index operation traces from the inside of a Postgresql DBMS by modifying related source codes. TPC-C benchmark [22] was used to create queries to the DBMS. We also compared a concurrent version of PIO B-tree with a concurrent B-tree that can operate in multi-threads with a fine-grained locking. We conducted the experiments on a Linux machine with 8-core CPU (2.0 GHZ), and 16GB main memory. We used three flashSSDs, Iodrive [5] a high-priced enterprise-class SSD, P300 [17] an enterprise-class SSD, and F120 [4] a consumer-class SSD.

B+-tree, BFTL, and PIO B-tree were implemented based on the description in their original papers. We converted the released code of FD-tree into a Linux-compatible one. For the concurrent B+-tree, we implemented B-link tree following the description in the paper [15]. The concurrent version of PIO B-tree was also implemented with a simple locking method. For every *speriod*, the entire OPQ is exclusively locked for the OPQ entries to be sorted. PIO B-tree exclusively locks the entire index for every OPQ flush operation. Since OPQ append operation can be instantly finished on main memory, and since OPQ sort and OPQ flush is not executed for every update but is periodically performed, even with this simple setting, search operations of PIO B-tree can be performed concurrently enough in multiple threads. The OPQ flush operation was always executed in a single thread due to the entire index lock.

### 4.1 Synthetic Workload

The indexes were initially built with 1 billion entries by using a bulk loader, thus occupying more than 8GB of storage space. The LRU buffer manager was employed for the indexes. The maximum available main memory space was fixed at 16MB. The parameters for PIO B-tree, *PioMax*, s*period, and bcnt* were fixed at 64, 5000, and 5000, respectively.

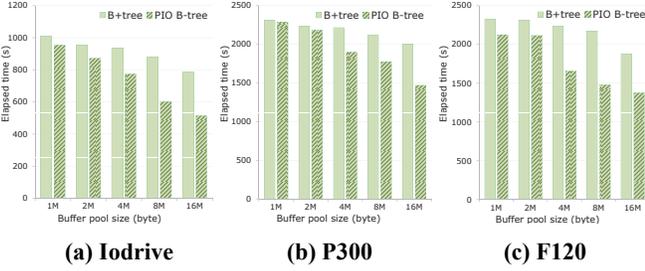

| (a) Iodrive | (b) P300 | (c) F120 |

Figure 9. Search time with different buffer sizes

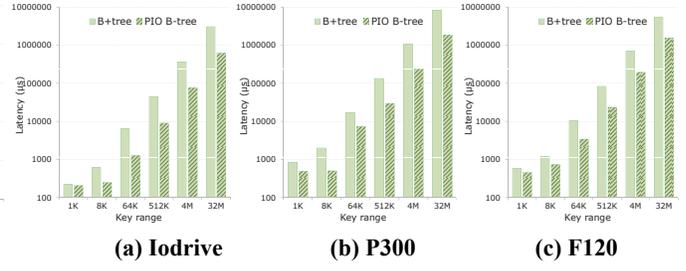

| (a) Iodrive | (b) P300 | (c) F120 |

Figure 10. Range search time with different ranges in log scale

### 4.1.1 B+-tree Node Size Optimization
This experimental set was configured with the search-only workload. Five million point-search operations were requested to B+-tree and PIO B-tree. We measured the total response time varying the buffer pool size from 1MB to 16MB. In order to choose the best node size for B+-tree, the utility/cost measure (3) was utilized. The leaf node size of the PIO B-tree index was configured according to (10).

As presented in Figure 9, compared to the B+-tree index, the PIO B-tree index has demonstrated better performance regardless of buffer pool sizes. The point-search operation of PIO B-tree was up to 1.5, 1.36, and 1.36 times faster than B+-tree on Iodrive, P300, and F120, respectively.

### 4.1.2 Parallel Range Search
We evaluated the range search performance by requesting queries having various key ranges. With the buffer pool size fixed at 16MB, the node size of B+-tree and PIO B-tree was configured the same as in Section 4.1.1. The gap between the start key and end key in the range was increased from 1024 (1K) to 33554432 (32M) by a factor of 8 at a time.

One hundred range searches were performed. Figure 10 represents the average elapsed time of a range search in log-scale with respect to the key ranges. PIO B-tree outperformed B+-tree with any given range on every flashSSD. The *prange* search of PIO B-tree was 5 times faster than the range search of B+-tree when the key range was greater than or equal to 64K on Iodrive. With the range greater than or equal to 512K, PIO B-tree was 4.5 times and 3.5 times faster than B+-tree on P300 and F120, respectively.

### 4.1.3 Update Operations
We evaluated the performance of *bupdate* by using update-only workloads. Since index-insert, index-delete, index-update operations demonstrated almost the same performance, we only report the insert workload result. The node size of PIO B-tree was configured the same as in Section 4.1.1. After allocating the main memory to OPQ and LSMap, the rest of main memory space was allocated to the buffer pool. We measured the elapsed time of five million insert operations, increasing the OPQ size from 1 page to (4095 – LSMap size) pages on a 4KB page basis. On the contrary, the buffer pool size is reduced from (4095-LSMap size) pages to 1 page. In order to assess the point-search performance degradation by the reduced buffer pool, we measured the elapsed time of five million point-search requests. Figure 11 presents this result.

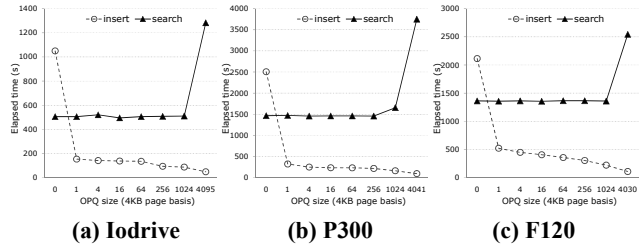

| (a) Iodrive | (b) P300 | (c) F120 |

Figure 11. Insert/search time of PIO B-tree with OPQ sizes

To compare the result with B+-tree, we measured insert and point-search performance of B+-tree with the same workloads. The B+-tree node size was configured the same as Section 4.1.1, and 16MB was allocated for the buffer pool. Compared to the B+-tree performance, only with the OPQ size of one (4KB), PIO B-tree has demonstrated remarkable insert performance. The PIO B-tree was 7.2, 8.2, and 4.3 times faster than B+-tree on Iodrive, p300, and F120, respectively. With a large OPQ size, it was up to 28 times faster than B+-tree (on P300 with OPQ size 4041). Note that PIO B-tree was 16.3 times faster than B+-tree in insert operations and at the same time 1.2 times faster than B+-tree in search operation, with the OPQ size of 1024 on P300.

### 4.1.4 Comparison with Flash-aware Indexes
We compared four indexes such as BFTL, B+-tree, FD-tree, and PIO B-tree in five workloads. The workloads were differentiated by the insert ratio and search ratio. We configured the workloads to have randomly chosen 10 million operations with the specified insert and search ratio. Since the flash-aware indexes take advantage of trade-offs between insert and search performance based on parameters, to be fair, we chose the best parameter according to the workload feature for each index. For this purpose, PIO B-tree was automatically configured according to Section 3.6, and other trees followed their own tuning methods. In the case of PIO B-tree, the buffer pool was configured as the remaining main memory space after allocating the OPQ and LSMap. In BFTL, the entire main memory space was consumed by its mapping table thus making no space left for the buffer pool.

As shown in Figure 12, the PIO B-tree was 2.5 to 13.7, 2 to 13, 2.1 to 15 times faster than the BFTL on Iodrive, P300, and F120, respectively. The PIO B-tree was 1.6 to 11, 1.4 to 10.9, and 1.6 to 10.5 times faster than the B+-tree on Iodrive, P300, and F120, while the PIO B-tree was 1.23 to 1.47, 1.24 to 1.45, and 1.24 to 1.46 times faster than the FD-tree on Iodrive, P300, and F120, respectively. In the graphs we differentiated insert and point-search time. The FD-tree's insert time was similar to that of PIO B-tree. The performance gap between PIO B-tree and FD-tree was mainly due to the PIO B-tree's faster point-search performance.

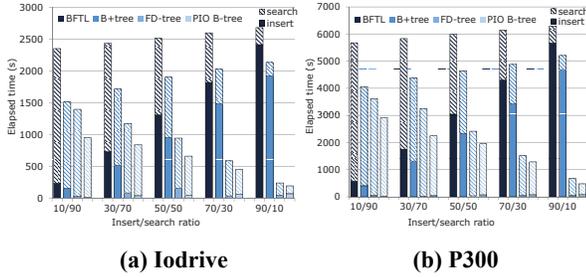
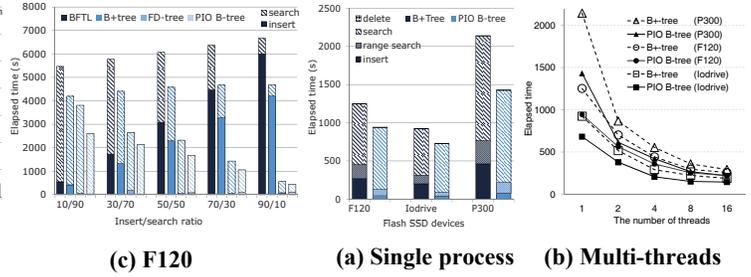

| (a) Iodrive | (b) P300 | (c) F120 | (a) Single process | (b) Multi-threads |

**Figure 12. Overall elapsed time in mixed workloads**   **Figure 13. Performance with TPC-C workload**

## 4.2 Index Trace from Postgresql (TPC-C)
TPC-C workload setting was configured as 100 warehouses and 100 clients. The total index relation size was approximately 1GB, configured with 8 index files for 8 index relations. The TPC-C workload showed higher temporal and spatial localities of index operations than synthetic workloads. The index trace in total of 10 million index operations consisted of 71.5 % point-search, 23.8 % inserts, 3.7% range searches, and 1% deletes.

First we compared the performance of PIO B-tree with B+-tree with main memory buffer size fixed at 4MB. In order to exclude the effects of node size optimization, we fixed the leaf node size of PIO B-tree at 1, and configured the node size of B+-tree and PIO B-tree with 4KB. The OPQ size was also fixed at 20. Other parameters (*bcnt*, *speriod*, *PioMax*) were fixed the same as synthetic workloads. As shown in Figure 13 (a), where delete time was so small that it is plotted as almost a dot, the PIO B-tree was 1.25 to 1.49 times faster than the B+-tree in total. For each operation, the PIO B-tree's insert, range search operation was 5.7 to 6.2, 1.9 to 2.1 times faster than insert, range search of the B+-tree, respectively. Next we compared the performance of concurrent PIO B-tree with concurrent B-tree (B-link tree) as increasing the number of parallel threads. As shown in Figure 13 (b), the PIO B-tree was 1.33 to 1.49, 1.3 to 1.47, and 1.17 to 1.32 times faster than the B-link tree on P300, Iodrive, and F120, respectively. The performance gap is due to the read/write interleaving, faster range search of PIO B-tree, and bandwidth drop in I/Os requested to a shared file of B-link tree. Since PIO B-tree has no dirty buffers, read/write operations are never interleaved until an OPQ flush operation is executed whereas the buffer manager employed in B-link tree causes frequent dirty buffer writes accompanied with buffer-miss reads. Even in the OPQ flush operation nearly no read/write interleaving occurs. Since 8 index files were configured, concurrent I/Os were distributed into different files, so the B-link tree performance was little affected by the bandwidth drop from the shared file.

## 5. RELATED WORK
Recently, intensive research has been conducted to reveal the internal architecture of flashSSDs [1, 2]. While the early studies paid less attention to the flashSSD features when outstanding I/Os are provided, Chen et al. [3] revealed that exploiting internal parallelism can significantly improve I/O performance. Recent other studies tried to improve flashSSD internal architecture design in order to provide more I/O parallelism inside flashSSDs [8, 21]. In the studies, channel-level parallelism was especially emphasized as a core of the flashSSD I/O parallelism. The features of package-level parallelism were also well studied in [11] that extracted flashSSD parameters through micro-benchmarks. Based on these, we confirmed it again that exploiting internal parallelism can significantly improve flashSSD bandwidth through benchmark tests on more various types of flashSSDs.

Unlike the previous studies that focused on uncovering internal parallelism features of flashSSDs and enforcing it inside flashSSDs, we focused on finding an efficient way to generate parallel I/Os in order to exploit the internal parallelism from outside of flashSSDs. By assessing different methods to create parallel I/Os, we suggest a new I/O request method (*psync* I/O). To the best of our knowledge, PIO B-tree is the first application that is designed for exploiting internal parallelism of flashSSDs.

Previous flash-aware B-trees [19, 23] focused on write optimization since their focus was enhancing B-tree performance on raw flash memory where the write latency is considerably higher than the read latency. As trade-offs, their search performance is degraded as much as the write-optimized level. FD-tree [16] is the first index structure that has a flashSSD-oriented design. The FD-tree index height is usually higher than B+-tree height since the fan-out of a FD-tree node is less than the fan-out of a B+-tree node. Therefore, the FD-tree's point-search performance is worse than the B+-tree's. PIO B-tree is the only flash-aware index that enhances update performance as well as point-search performance of B+-tree.

## 6. CONCLUSIONS
Due to the embedded flash memory packages, the internal parallelism is an inherent feature of flashSSDs. In this paper, we found an efficient way (*psync* I/O) to generate parallel I/Os for exploiting the internal parallelism. We presented PIO B-tree that optimizes B+-tree in the node size and index algorithms. In the experiments, PIO B-tree has demonstrated up to 5 times enhanced performance in the range search. PIO B-tree enhanced B+-tree's insert performance by a factor of up to 16.3, enhancing the search time by a factor of 1.2 at the same time. Moreover, PIO B-tree outperformed other flash-aware indexes in various workloads.

## 7. ACKNOWLEDGMENTS

We thank Prof. Bongki Moon at University of Arizona for a helpful discussion about async I/O; Yinan and coauthors of [16] for sharing the source codes and replies to our queries; anonymous reviewers for insightful comments. This paper was supported by Basic Science Research Program through the National Research Foundation of Korea (NRF) funded by the Ministry of Education, Science and Technology (2011-0004382).

## APPENDIX

Cost models with a buffer pool are based on (5) and (7). We assume that the buffer manager caches all the nodes, starting from the root level until it reaches the *LastLevel* where no more room for the buffered node is left. $H_b$ denotes the buffered height, the height of a sub-tree including nodes from the root level to the *LastLevel*, whereas $H_{nb}$ denotes the non-buffered height. *Cvrg* denotes the coverage of the buffered nodes at *LastLevel* (i.e. the ratio of the buffered node number to the non-buffered node number). Since the buffered node read-cost is negligible, the remainder is the time to read nodes of the non-buffered levels and the partially buffered level (*LastLevel*). The cost for the non-buffered level is reading as many nodes as non-buffered height $H_{nb}$. For the partially buffered level, since the last level includes buffered nodes, the cost for reading a node in this level is less than $P_r$. The probability of reading non-buffered nodes is $(1 - Cvrg)$, assuming node key values in a uniform distribution. The average cost for a B+-tree index operation can be represented as (11).

$$C'_{b+} = (H_{nb} + (1 - Cvrg)) \cdot P_r + R_i \cdot P_w \quad (11)$$

In order to induce $H_{nb}$ and *Cvrg*, we formulate (12) to (14).

$$M = \sum_{i=0}^{LastLevel} F'^i = \frac{1 - F'^{LastLevel+1}}{1 - F'}$$

$$LastLevel \cong \log_{F'}\{(F'-1) \cdot M\} - 1 \cong \log_{F'} M$$

$$H_b = LastLevel + 1 \cong \log_{F'} M + 1 \quad (12)$$

$$H_{nb} = \lfloor H - H_b \rfloor = \lfloor \log_{F'}(N/M) - 1 \rfloor \quad (13)$$

The coverage is the ratio of the number of nodes buffered at *LastLevel* to the total number of nodes at *LastLevel*.

$$Cvrg = F'^{H_b - 1} / F'^{H - H_{nb} - 1} = 1/F'^{(H-H_b)\%1} \quad (14)$$

The formula (11) is completed by substituting $H_{nb}$ and $Cvrg$ with (13) and (14), and letting $\eta$ substitute $\log_{F'}(N/M) - 1$.

The cost model for PIO B-tree can be induced in a similar manner based on (7). First, with the buffer pool considered the node read cost of an update operation can be represented as (15). Since the same node is read only once, the node read cost for non-buffered height becomes $\sum_{\ell=\lceil\eta\rceil}^{H-1}\left(\frac{1}{G(\ell)}\right) \cdot P'_r$, and the node read cost for *LastLevel* becomes $\frac{1/F'^{(\eta\%1)}}{G(\log_{F'}(M-O)-1)} \cdot P'_r$.

$$\left(\sum_{\ell=\lceil\eta\rceil}^{H-1}\left(\frac{1}{G(\ell)}\right) - \frac{1/F'^{(\eta\%1)}}{G(\log_{F'}(M-O)-1)}\right) \cdot P'_r \quad (15)$$

Second, the available main memory space is reduced as $(M - O)$. Since OPQ occupies a portion of the main memory. By applying theses changes to (7), $C'_{pio}$ can be induced as (9).